# Observation of Umklapp processes in non-crystalline materials

Tullio Scopigno*, Matteo D'astuto[+], Michael Krisch[+], Francesco Sette[+] and Giancarlo Ruocco[#].

*Dipartimento di Fisica and INFM, Universita' di Trento, -xxxxx, Trento, Italy.
[+]European Synchrotron Radiation Facility, B.P. 220 F-38043 Grenoble, France.
[#]Dipartimento di Fisica and INFM, Universita' di L'Aquila, I-67100, L'Aquila, Italy.

ABSTRACT

*Umklapp processes are known to exist in cristalline materials, where they control important properties such as thermal conductivity, heat capacity and electrical conductivity. In this work we report the provocative observation of Umklapp processes in a non-periodical system, namely liquid Lithium. The lack of a well defined periodicity seems then not to prevent the existence of these scattering processes mechanisms provided that the local order of the systems i.e. the maxima of the static structure factor supply the equivalent of a reciprocal lattice vector in the case of cristalline materials.*

Umklapp processes (U-processes) are scattering processes among elementary excitations in crystals where *momentum conservation* is fulfilled with a contribution from the lattice (*1*). Their existence has fundamental consequences on transport properties. More specifically, U-processes of phonons in crystals determine the finite thermal conductivity, and are involved in the electron-phonon coupling phenomena with implication on the electrical resistivity and the electronic contribution to the heat capacity. These U-processes are understood in crystals, where they are associated with the lattice periodicity. Their existence in a topologically disordered system is matter of debate - They are unexpected because of the absence of *long range* order, but, one could also imagine that just *local* order could support them, at least for those phonon-like excitations observed in disorder systems at wavelengths comparable to the inter-particle separation.

In this Letter, we report the observation of U-like-processes in liquid lithium. Using Inelastic X-ray Scattering (IXS), we observe that the phonon-like acoustic excitations of this topologically disordered system are 'reflected' by the quasi-periodicity responsible for the First Sharp Diffraction Peak (FSDP) in the static structure factor.

Transport properties in a material determine the flow of characteristic quasi-particles induced by an external field (temperature gradient, electric field, etc.). In an ideal infinite perfect crystal, with no lattice imperfections, impurities and surface effects, the existence of a stationary flow is a consequence of interactions among quasi-particles and of U-processes. The interaction is necessary to exchange energy and momentum, and, in particular to allow scattering processes as $\boldsymbol{k_1},\boldsymbol{k_2} \rightarrow \boldsymbol{k_3}$, where $\hbar\boldsymbol{k}$ is the quasi-particle momentum. However, as the total momentum is conserved ($\boldsymbol{k_1}+\boldsymbol{k_2}=\boldsymbol{k_3}$), the interaction alone is not sufficient to reach equilibrium. This is possible thanks to the U-processes. Their fundamental role is to open further scattering channels where the momentum of the quasi-particles in the initial state is different from that of the final state. The difference, necessary for the *total* momentum conservation in the scattering event, is supplied by the crystal periodicity via the reciprocal lattice vectors $\boldsymbol{G}$: $\boldsymbol{k_1}+\boldsymbol{k_2}=\boldsymbol{k_3}+\boldsymbol{G}$. Therefore, it is evident that the concept of U-process is intimately related to the existence of a periodic lattice, via the vectors $\boldsymbol{G}$, and to the periodic repetition of the crystal quasi-particles dispersion relation in successive Brillouin zones.

In *crystals*, particularly important excitations are the phonons. Phonon U-processes are, for example, directly responsible for the thermal conductivity in insulators. It is firmly believed that also in *disordered systems* the thermal transport properties are determined by the excitation spectrum of the density fluctuations (*2*). In particular, in glasses, the effect of the disorder on the collective dynamics is supposed to be responsible for the observed anomalies in the heat capacity and thermal conductivity behavior when compared to their crystalline counterpart. In this respect, recent experimental (*3-8*) and numerical (*9-13*) works have demonstrated that, in glasses and disordered systems in general, propagating acoustic like collective excitations exist up to a momentum ($\hbar Q$) region approaching the $Q$ value of the FSDP, $G_{FDSP}$. These excitations, with increasing $Q$, show clear effects due to the disorder as, for example, the increasing non-plane wave character of their spatial eigenfunctions. Nevertheless, one finds for these excitations a dispersion relation $E(Q)$ vs. $Q$ strikingly similar to that of crystals. Among others, a linear behavior of $E(Q)$ at small $Q$, and a bend down of $E(Q)$ evolving, in certain cases where these excitations are particularly well defined, into a maximum around $G_{FSDP}/2$. The behavior of the excitations dispersion relation in disordered materials, with its similarity to the crystalline case, induces to speculate that also in these systems is possible to define a pseudo-Brillouin Zone, whose first replica covers the region $0 < Q < G_{FSDP}/2$. Pushing further this speculation into higher order Brillouin zones, one may expect that U-like processes also exist in disordered systems.

The previous concepts are pictorially illustrated in Fig. 1 for the one-dimensional case. Here, the dashed line sketches an acoustic phonon dispersion relation $E(Q)$ in a perfect chain with lattice constant $a$. This dispersion relation periodically repeats itself with a period $G=2\pi/a$. In the same plot, we also sketch the static structure factor $S(Q)$, which in such a periodic system corresponds to Bragg peaks at $Q$ values $G_n=2\pi n/a$ (full vertical line). Let's imagine now performing the following scattering experiment in such a perfect chain: as shown by the dash-dotted line in Fig 1a, we scan the exchanged momentum at the fixed energy $E_0$. We will observe a sharp peak, the Brillouin peak, at $Q = Q_0 \approx \hbar E_0/v$, where $v$ is the sound velocity and $Q$ is low enough to be in the linear dispersion region. This peak will also be observed at $Q = G \pm Q_0$, where $G$ is any of the reciprocal lattice vectors. The presence of these others peaks, the Umklapp peaks, is due to the fact that the phonon momentum is defined to within a reciprocal lattice vector. Naively, one can says that the phonon of momentum $Q_0$ is "reflected" by the lattice (or by the Bragg peak at $Q = G$) and its signature in the spectrum of the reflected phonons is the appearance of peaks, other than at $Q_0$, on the two sides of the Bragg peak, i. e. at $G \pm Q_0$. In Fig. 1b, we report an attempt to translate such scattering experiment from a periodic array into a disordered chain. Supposing for the time being that the excitations in such "glassy" system were phonon-like plane waves, one still observes a sharp Brillouin peak at $Q = Q_0$. However, absence of periodicity destroys the sharp Bragg peaks, and the static structure factor is characterized by a feature, the so called First Sharp Diffraction Peak (FSDP), centered at $Q = G_{FSDP} \cong 2\pi/a$ ($a$ is the average near neighbor distance). This feature has a finite width, $\Delta G_{FSDP}$, determined by the distribution of nearest-neighbors distance that typically is an important fraction of $G_{FSDP}$. In analogy to the crystal case, one may expect that this FSDP is also capable to "reflect" the phonons. However, with respect to the peak at $Q_0$, the peaks at $Q_0= G_{FSDP} \pm Q_0$ will have a width increased by $\Delta G_{FSDP}$. In this scenario, the initial working hypothesis that the excitation at $Q = Q_0$ is a plane wave is not really necessary. In fact, taking into account that in a glass the disorder induces a width of the Brillouin line, one expects that also the Umklapp peaks widths will increase accordingly. Finally, one expects that the simple description sketched in Fig. 1b, although more complicated, is still valid in a real three-dimensional system, at least for longitudinal phonons on a qualitative ground. A quantitative description has been derived by Carpenter and Pellizzari (*14*), under the assumption of pure plane waves excitations, and further refined by Ribeiro, Wilson and Madden (*13*).

Scattering experiments as the one sketched in Fig. 1b have been performed in disordered systems (*15,16*) using Inelastic Neutron Scattering (INS). However, no experiment so far has been able to cover the whole interesting $Q$-$E$ region, going at least from $Q_0$ and up to the FSDP. Moreover, in the existing data, no attempts have been made to identify the existence of U-processes.

Liquid lithium is a disordered system that has been already thoroughly investigated using the IXS technique (*17*). Compared to prototypical glassy systems (vitreous silica for example) it shows a particularly favorable contrast between inelastic and elastic signals. This property makes liquid lithium a very good candidate to search for Umklapp features in the Q region around $G_{FSDP}$, where the elastic contribution to the spectra has its maximum intensity and where the first Umklapp peaks are expected to be.

The IXS experiment has been carried out at the very high-energy resolution IXS beam-line (ID28) at the European Synchrotron Radiation Facility. The instrument consists of a back-scattering monochromator and five independent analyser systems, held one next to each other with a constant angular offset on a 7~m long analyser arm. We utilised the Si(9 9 9) configuration, giving a total instrumental energy resolution of 3.0 meV full-width-half-maximum (fwhm) (*18*), and an offset of 2.4 nm$^{-1}$ between two neighbour analysers. The spectra have been measured as a function of $Q$ transfer and at different constant $E$ transfers by setting a constant temperature difference between the monochromator and analysers crystals and by rotating the analyser arm. The measured Q range is 1-55 nm$^{-1}$, and the Q resolution is 0.3 nm$^{-1}$ fwhm. The spectra were normalised to the incoming photon flux, and, for improved statistical accuracy, the signals from different analysers in overlapping Q regions was averaged together. Each scan took about 150 minutes, and each spectrum was obtained by the typical average of five scans. The liquid lithium sample was in an uncapped container made of austenitic stainless steel. A resistance heater was used to keep the liquid at 475 K, i.e. slightly above the melting point at 453 K. The 20 mm long sample was loaded in an argon glove box and kept in a 10$^{-6}$ mbar vacuum. Empty vacuum chamber measurements were also performed. They gave either the flat electronic detector background of 0.6 counts/min at $Q>8$ nm$^{-1}$, or, for $Q<8$ nm$^{-1}$, a small elastic signal due to the vacuum chamber Kapton windows (50 µm thick), whose intensity gives a negligible contribution to the inelastic spectra reported here.

The IXS spectra measured at the constant $E$ values of 15, 20, 25, 30, 35, and 40 meV are reported in Fig. 2 (full circles). In the same figure the static structure $S(Q)$ is also shown as a full line, and the FSDP is found at $G_{FSDP} \approx 25$ nm$^{-1}$ with a width of $\Delta G_{FSDP} \approx 4$ nm$^{-1}$. The value of $G_{FSDP} = 2\pi/a$ is consistent with the average nearest neighbour distance $a \approx 0.24$ nm.

The inelastic spectra, $I(Q,E)$, reported in Fig. 2, normalised only to the incident photon flux, could be reduced to the dynamic structure factor $S(Q,E)$ by correcting for the lithium atomic scattering form factor, $f(Q)$, and for a coefficient, $g(Q)$, that takes into account the $Q$-dependent response function of the IXS spectrometer: $S(Q,E) = I(Q,E) / [ f(Q) g(Q) ]$. Both $f(Q)$ and $g(Q)$ are smooth functions which decrease monotonically with increasing $Q$. However, independently from the specific knowledge of the functions $f(Q)$ and $g(Q)$, the spectra $I(Q,E)$ can be put on an absolute scale – thus deriving the $S(Q,E)$ - using the well known sum rules of the dynamic structure factor (*19*). This procedure has been exploited elsewhere (*17*). In the present work we prefer to report the raw data to emphasise the direct observation of the inelastic features without any manipulation. In Fig. 2 the spectra have been normalised to an arbitrary value such that their intensity is comparable to that of the $S(Q)$ in the high $Q$ region.

The spectra in Fig. 2 show in the small $Q$ region a Brillouin peak that increases its $Q$ position and width with increasing energy. More importantly, the spectra show also other inelastic peaks at higher Q values (marked by the arrows). This ensemble of inelastic features can be interpreted within the same framework as the one sketched in Fig. 1b. In fact, one observes, at low energy transfer $E$, not only the Brillouin peak at $Q_0(E)$, but also a second peak at approximately $G_{FSDP} - Q_0(E)$. Moreover, one clearly observes that, with increasing $E$, the peaks at $Q_0(E)$ and $G_{FSDP} - Q_0(E)$ get closer and, at the energy $E \approx 40$ meV, they merge together at a $Q$ value correspondent to $G_{FSDP} / 2$. This behaviour, as previously emphasised in Fig. 1b, is precisely the one expected in presence of Umklapp peaks. These data, therefore, provide a compelling evidence that the U-processes are active also in disordered systems and show that the effect of disorder can be framed within the qualitative model presented in Fig. 1b. In this respect, in fact, the Umklapp peak at $G_{FSDP} - Q_0$ is considerably broader than the Brillouin peak at $Q_0$. This broadening is also responsible for the even more reduced visibility of the Umklapp process in the third pseudo-Brillouin zone that is expected at $G_{FSDP} + Q_0$. We notice however that this second family of Umklapp peaks at $G_{FSDP} + Q_0$ is nevertheless observable in the limit of large $E$ as emphasised by the dashed arrows.

In conclusion we have shown that in a system without any periodic order as a monatomic liquid, one observes inelastic excitations that can be interpreted as the non-crystalline counterpart of Umklapp peaks. These peaks are no-longer sustained by the periodicity of the lattice –as in crystals- but are due to the "reflection" of the wave-like (Brillouin) excitations from the short range order that still exists in the disordered material. Contrary to the crystalline case, however, the presence of disorder is responsible for an increasingly large broadening of these Umklapp peaks with increasing momentum. As in crystal, on the other hand, the present observation allows to speculate that the Umklapp processes play an important role in the transport properties of disordered materials.

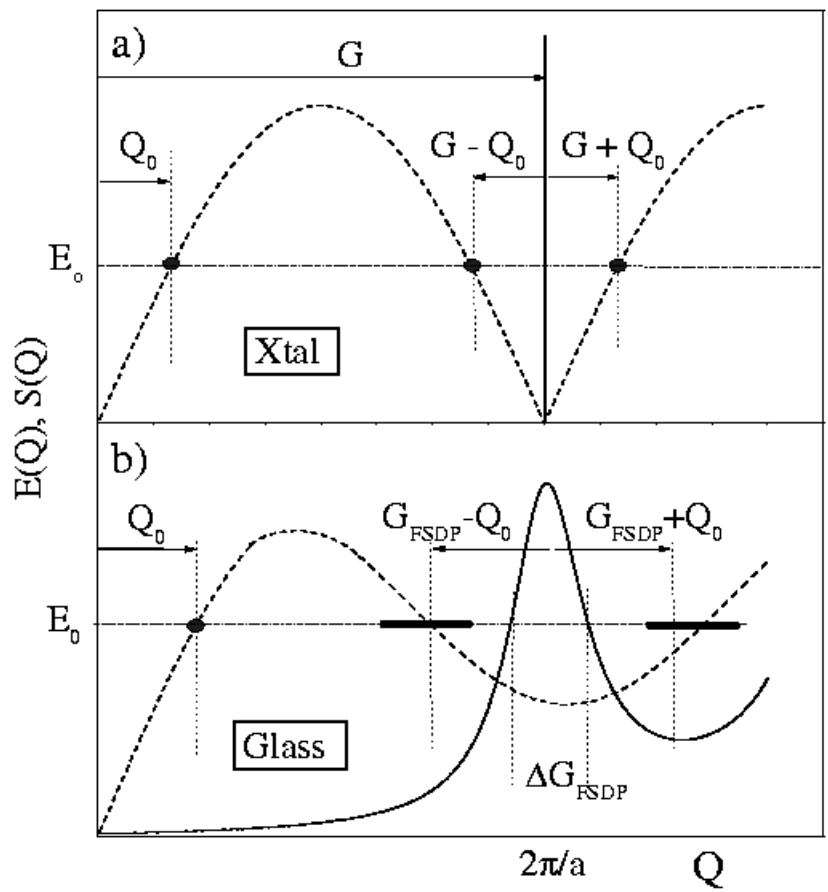

Fig. 1 - T. Scopigno et al. - "Umklapp processes in ..."

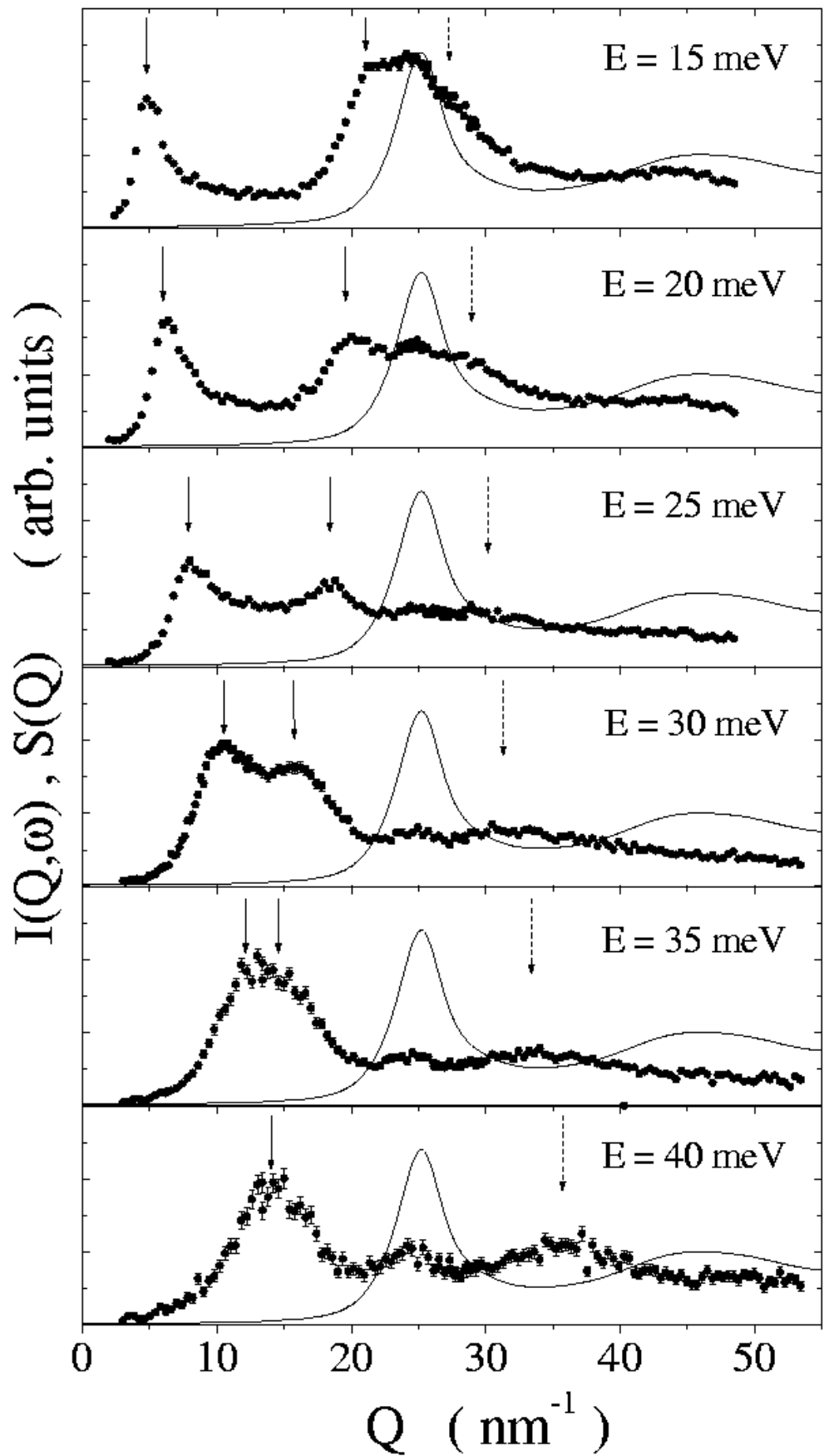

Fig. 2 - T. Scopigno et al. - "Umklapp processes in ..."

FIGURE LEGENDS

1 – **Sketch of the dispersion relation** $E(Q)$ (dashed lines) and of the static structure factor $S(Q)$ (full lines) of a one dimensional system. Fig. 1a reports the case of an ordered chain; Fig. 1b that of a disordered chain.

2 – **IXS spectra of liquid lithium** at T=475 K (full circles) taken at the indicated constant energy as a function of $Q$. The arrows mark the position of the main spectral features (Brillouin peaks at low $Q$ and Umklapp peaks around the FSDP). The full line in each panels is the $S(Q)$ reported on an arbitrary intensity scale.